\documentclass[aps,prb,twocolumn,preprintnumbers,amsmath,amssymb,superscriptaddress]{revtex4}%

\usepackage{graphicx}%
\usepackage{dcolumn}
\usepackage{amsmath}
\usepackage{color}
\usepackage{multirow}

\begin{document}

\title{Anisotropy induced vortex lattice rearrangement in CaKFe$_4$As$_4$}
\author{Rustem~Khasanov}
 \email{rustem.khasanov@psi.ch}
 \affiliation{Laboratory for Muon Spin Spectroscopy, Paul Scherrer Institut, CH-5232 Villigen PSI, Switzerland}
\author{William R. Meier}
 \affiliation{Division of Materials Science and Engineering, Ames Laboratory, Ames, Iowa 50011, USA}
 \affiliation{Department of Physics and Astronomy, Iowa State University, Ames, Iowa 50011, USA}
\author{Sergey L. Bud'ko}
 \affiliation{Division of Materials Science and Engineering, Ames Laboratory, Ames, Iowa 50011, USA}
 \affiliation{Department of Physics and Astronomy, Iowa State University, Ames, Iowa 50011, USA}
\author{Hubertus Luetkens}
 \affiliation{Laboratory for Muon Spin Spectroscopy, Paul Scherrer Institut, CH-5232 Villigen PSI, Switzerland}
\author{Paul C. Canfield}
 \affiliation{Division of Materials Science and Engineering, Ames Laboratory, Ames, Iowa 50011, USA}
 \affiliation{Department of Physics and Astronomy, Iowa State University, Ames, Iowa 50011, USA}
\author{Alex Amato}
 \affiliation{Laboratory for Muon Spin Spectroscopy, Paul Scherrer Institut, CH-5232 Villigen PSI, Switzerland}

\begin{abstract}
The magnetic penetration depth anisotropy $\gamma_\lambda=\lambda_{c}/\lambda_{ab}$ ($\lambda_{ab}$ and $\lambda_{c}$ are the in-plane and the out-of-plane components of the magnetic penetration depth) in a CaKFe$_4$As$_4$ single crystal sample (the critical temperature $T_{\rm c}\simeq 35$~K) was studied by means of muon-spin rotation ($\mu$SR). $\gamma_\lambda$ is almost temperature independent for $T\lesssim 20$~K ($\gamma_\lambda\simeq 1.9$) and it reaches $\simeq 3.0$ by approaching $T_{\rm c}$. The change of $\gamma_\lambda$ induces the corresponding rearrangement of the flux line lattice (FLL), which is clearly detected via enhanced distortions of the FLL $\mu$SR response. Comparison of $\gamma_\lambda$ with the anisotropy of the upper critical field ($\gamma_{H_{\rm c2}}$) studied in Phys. Rev B {\bf 94}, 064501 (2016), reveals that $\gamma_\lambda$ is systematically higher than $\gamma_{H_{\rm c2}}$ at low-temperatures and approaches $\gamma_{H_{\rm c2}}$ for $T \rightarrow T_{\rm c}$. The anisotropic properties of $\lambda$ are explained by the multi-gap nature of superconductivity in CaKFe$_4$As$_4$ and  are caused by anisotropic contributions of various bands to the in-plane and the out-of-plane components of the superfluid density.
\end{abstract}


\maketitle

In the majority of superconducting compounds discovered so far, the crystal structure, as well as the electronic and phononic band
structures are all far from being isotropic.
Anisotropic superconductors are usually treated within the phenomenological anisotropic Ginzburg-Landau (AGL) theory,\cite{Ginzburg_ZhETPh_1950, Ginzburg_ZhETPh_1952} which follows from the isotropic Ginzburg-Landau (GL) approach via replacement of the effective mass $m^\ast$ in the GL free energy functional by an effective mass tensor, with values $m_{a}^\ast$, $m_{b}^\ast$, and $m_{c}^\ast$ along the principal $a-$, $b-$, and $c-$axes.\cite{Caroli_PKM_1963, Thiemann_PRB_1989} In the most usual case of uniaxial anisotropy, all anisotropies are incorporated into the single parameter:\cite{Tinkham_75}
\begin{equation}
\gamma=\;\gamma_{\lambda}\equiv \frac{\lambda_{c}}{\lambda_{ab}}=\sqrt{\frac{m_c^\ast}{m_{ab}^\ast}}=\; \gamma_{H_{\rm c2}}\equiv \frac{H_{\rm c2}^{\parallel ab}}{H_{\rm c2}^{\parallel c}}=\frac{\xi_{ab}}{\xi_c}.
 \label{eq:gamma}
\end{equation}
Here $\gamma_\lambda$ and $\gamma_{H_{\rm c2}}$ are the anisotropy of the magnetic penetration depth ($\lambda$) and the upper critical field ($H_{\rm c2}$), respectively, and $\xi$ is the coherence length.
In AGL theory the same effective mass tensor determines the anisotropy of $\lambda$ and $H_{\rm c2}$, thus making both $\gamma_\lambda$ and $\gamma_{H_{\rm c2}}$ temperature and field independent.\cite{Gorkov_ZhETPh_1964}

Note, however, that the GL theory is strictly valid only for $T\rightarrow T_c$. Following Kogan,\cite{Kogan_PRL_2002} away from $T_{\rm c}$ the theoretical approach for calculating $H_{\rm c2}$ (the position of the second
order phase transition in high fields) has little in common with evaluation of $\lambda$, so that the anisotropies $\gamma_\lambda$ and $\gamma_{\rm H_{\rm c2}}$ are not the same and can be substantially different. In MgB$_2$, {\it e.g.}, both anisotropies approach a common value $\gamma_\lambda=\gamma_{H_{\rm c2}}\simeq 1.75$ at $T_{\rm c}$ and become $\gamma_\lambda\simeq 1.2$ and $\gamma_{H_{\rm c2}}\simeq6$ at low temperatures.\cite{Angst_PRL_2002, Angst_Book_2004, Fletcher_PRL_2005} Different $T$ dependencies of $\gamma_\lambda$ and $\gamma_{\rm H_{\rm c2}}$ have also been reported for various cuprate and Fe-based superconductors.\cite{Khasanov_JSNM_2008, Khasanov_PRL_2007, Prozorov_PhysC_2009, Khasanov_PRL_2009, Khasanov_PRL_2009_2, Weyeneth_JSNM_2009, Weyeneth_JSNM_2009_2, Bendele_PRB_2010, Katterwe_PRB_2014}

\begin{figure}[htb]
\includegraphics[width=1.0\linewidth]{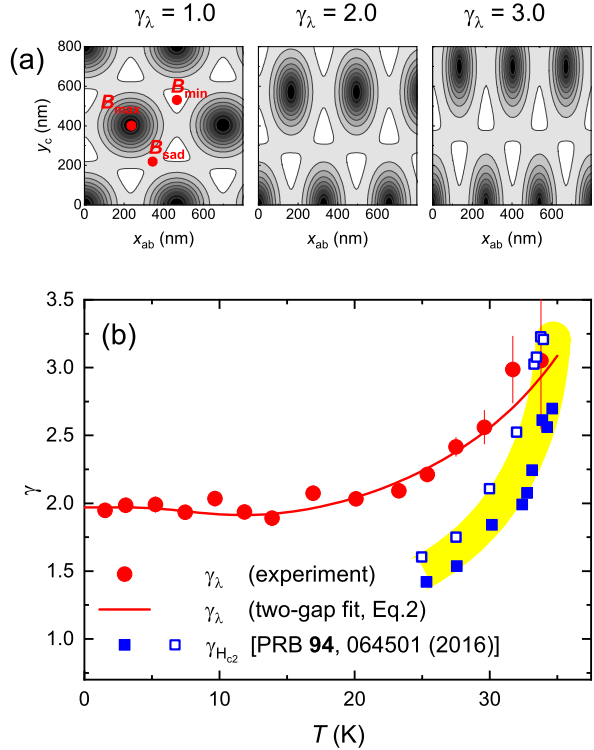}
 \vspace{-1cm}
\caption{(a) The contour plot of the field variation within the triangular vortex lattice of anisotropic superconductor with $B_{\rm ex} \parallel ab$   ($B_{\rm ex}=11$~mT, $\gamma_\lambda=\lambda_c/\lambda_{ab}=1.0$, 2.0, and 3.0; see the Supplemental part, Ref.~\onlinecite{Supplemental_Material}). $B_{\rm min}$,
$B_{\rm max}$, and $B_{\rm sad}$ are the minimum, maximum and the saddle point fields. (b) The anisotropies $\gamma_\lambda$ and $\gamma_{H_{\rm c2}}$ of CaKFe$_4$As$_4$. Closed circles are $\gamma_\lambda$ points from the present study.  The closed and open squares correspond to $\gamma_{H_{\rm c2}}(T)$ from Ref.~\onlinecite{Meier_PRB_2016}, as obtained by using the 'onset' and 'offset' criteria of $T_{\rm c}$ determination, respectively. The solid red line is the fit of two-gap model (Eq.~\ref{eq:two-gap}) to $\lambda_{ab}^{-2}(T)$ and $\lambda_c^{-2}(T)$ data.}
 \label{fig:Gamma}
\end{figure}

It is worth to mention here, that the change of $\gamma_\lambda$ would necessarily lead to a corresponding flux line lattice (FLL) rearrangement. As follows from Fig.~\ref{fig:Gamma}~(a), for the external field $B_{\rm ex}$ applied along the $ab-$plane ($B_{\rm ex}\parallel \; ab$),  the increase of $\gamma_\lambda$ shortens the distance between vortices within the $ab$ plane and enhances the intervortex distances in the $c-$direction. The displacement of vortex lines could be quite big. The simple estimate reveals that the increase of $\gamma_\lambda$ from $ 1.0$ to $3$  leads to decrease/increase of the in-plane/out-of-plane intervortex distances by almost half of the isotropic FLL unit cell [see Fig.~\ref{fig:Gamma}~(a)]. This effect can be studied, {\it e.g.} by using techniques visualizing the vortex distribution in superconducting materials, such as small-angle neutron scattering (SANS), tunneling or magnetic decoration. The important limitation comes, however, from the sample size and morphology. Most of the available good quality single crystals are very thin along the crystallographic $c-$direction, as well as a good surface preparation (needed for magnetic decoration and tunneling experiments) in $B_{\rm ex} \perp c$ orientation is challenging.

In this paper, we report on measurements of the magnetic penetration depth anisotropy $\gamma_\lambda=\lambda_{c}/\lambda_{ab}$ in a CaKFe$_4$As$_4$ single crystal sample ($T_{\rm c}\simeq 35$~K) by means of the muon-spin rotation ($\mu$SR). Comparison of $\gamma_\lambda$ with $\gamma_{H_{\rm c2}}$ studied in Ref.~\onlinecite{Meier_PRB_2016} shows that $\gamma_\lambda$  is higher than $\gamma_{H_{\rm c2}}$ all the way up to $T_{\rm c}$. Only in the narrow region close $T_{\rm c}$, $\gamma_\lambda=\gamma_{H_{\rm c2}}\simeq 3.0$ [Fig.~\ref{fig:Gamma}~(b)]. The change of $\gamma_\lambda$ with temperature induces the corresponding rearrangement of the flux line order, which was detected via the enhanced distortions of FLL $\mu$SR response. The temperature dependencies of the in-plane and the out-of-plane components of the superfluid density ($\lambda_{ab}^{-2}$ and $\lambda_{c}^{-2}$) were found to be well described within the two-gap scenario, thus suggesting that the multiple-band nature of superconductivity in CaKFe$_4$As$_4$ is well pronounced for both $c-$ and $ab-$ directions.

CaKFe$_4$As$_4$ single crystal with dimensions of $\simeq 4.0 \; \times \; 4.0 \; \times \; 0.1$~mm$^3$ was grown from a high-temperature Fe-As rich melt,\cite{Meier_PRB_2016, Meier_PRM_2017} and it is characterized via magnetization measurements (see the Supplemental part, Ref.~\onlinecite{Supplemental_Material}).  The $\mu$SR measurement were carried out at the $\pi$M3 beam line using the GPS spectrometer (Paul Scherrer Institute, Switzerland).\cite{Amato_RSI_2017} The zero-field (ZF) and transverse-field (TF) $\mu$SR measurements were performed at temperatures from $\simeq$1.5 to 50~K. In two sets of TF-$\mu$SR experiments the external magnetic field ($B_{\rm ex}\simeq 11$~mT) was applied parallel to the $ab-$plane and the crystallographic $c-$axis of the crystal, respectively. A fraction of TF-$\mu$SR data for $B_{\rm ex} \parallel c$ was previously reported in Ref.~\onlinecite{Khasanov_PRB_2018}. ZF-$\mu$SR data are discused in the Supplemental part, Ref.~\onlinecite{Supplemental_Material}. A special sample holder  designed to measure thin samples by means of $\mu$SR was used.\cite{Khasanov_PRB_2016} The experimental data were analyzed using the MUSRFIT package.\cite{MUSRFIT}

\begin{figure}[htb]
\includegraphics[width=1\linewidth]{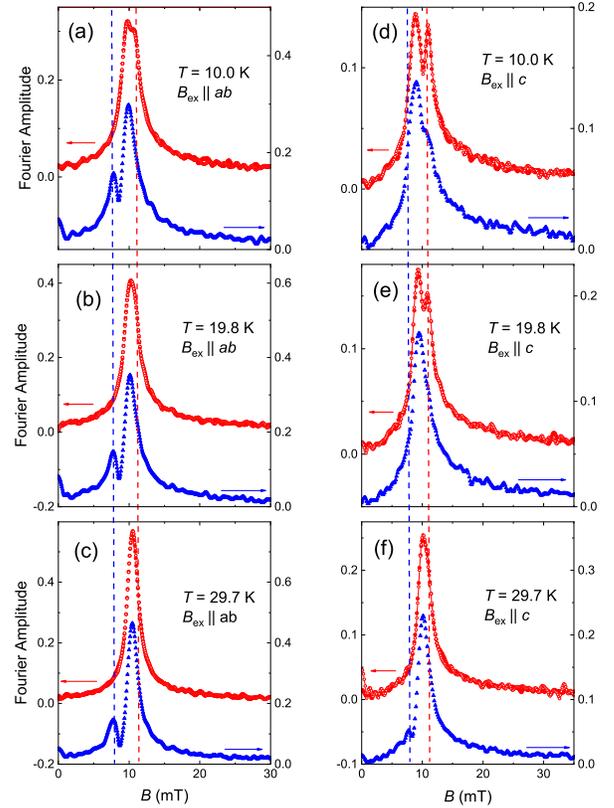}
 \vspace{-1cm}
\caption{The Fast Fourier transform  of the TF-$\mu$SR time spectra. Panels (a)--(c)  correspond to $B_{\rm ex}\parallel ab$ and panels (e)--(f) to $B_{\rm ex}\parallel c$ set of measurements, respectively.\cite{comment} The solid and open red symbols are $P(B)$'s obtained by following FCW and FCC protocols at $B_{\rm ex}\simeq 10.8$~mT (see text for details). The blue closed symbols are  $P(B)$'s after the field shift down to $\simeq 8$~mT. Dashed lines correspond to the applied field (10.8~mT) and to the 'shifted' field (8.0~mT), respectively.}
 \label{fig:Field-Shift}
\end{figure}

The homogeneity of the superconducting state in CaKFe$_4$As$_4$ was checked by performing series of field-shift experiments. Figure~\ref{fig:Field-Shift} exhibits the Fast Fourier transform  of the TF-$\mu$SR time spectra, which reflects the internal field distribution $P(B)$. The panels (a) to (c)  corresponds to $B_{\rm ex}\parallel \; ab$ and panels (e) to (f) to $B_{\rm ex}\parallel c$ set of measurements, respectively. The solid red symbols are $P(B)$'s obtained after cooling the sample at $B_{\rm ex}=10.8$~mT from a temperature above $T_c$ down to 1.48~K and subsequent warming it up to the measurement temperature [field-cooled warming (FCW) procedure]. The open red symbols are $P(B)$'s obtained after direct cooling from $T>T_{\rm c}$ to the measurement temperature [field-cooled cooling (FCC) procedure]. The blue closed symbols are  $P(B)$'s distributions collected by following the FCW protocol and subsequent decrease of the external field down to $\simeq 8$~mT (field-shift procedure). From the data presented in Fig.~\ref{fig:Field-Shift} the following three important points emerges.
(i) The main part of the signal, accounting for approximately 90\% of the total signal amplitude, remains unchanged within the experimental error after a field shift. Only the sharp peak ($\simeq10$\% of the signal amplitude) follows exactly the applied field. It is attributed, therefore, to the residual background signal from muons missing the sample (see also Refs.~\onlinecite{Sonier_PRL_1994, Khasanov_PRB_2016}).
(ii) The asymmetric $P(B)$ distributions shown in Fig.~\ref{fig:Field-Shift} possess the basic features expected for a well aligned vortex lattice, {\it i.e.}, the cutoff at low fields ($B_{\rm min}$), the peak arising from the saddle point midway between two adjacent vortices ($B_{\rm sad}$), and a long tail towards high fields caused by regions around the vortex core ($B_{\rm max}$).\cite{Maisuradze_JPCM_2009,Khasanov_PRB_2018} Note that the definition of "well aligned" FLL is different, {\it e.g}, for $\mu$SR and SANS experiments. In $\mu$SR case, due to the microscopic nature of  $\mu$SR technique, the field at the muon stopping site is determined by the {\it local} flux line arrangement (within a couple of FLL unit cells).  SANS experiment collects reflections from flux-line planes, which requires a {\it long range} arrangement of FLL. Following scanning tunneling microscopy (STM) results, the FLL in CaKFe$_4$As$_4$ is well arranged on the short length scale, while the long range arrangement is missing [see Figs.~3~(d), (e), and (f) in Ref.~\onlinecite{Fente_PRB_2018}].
(iii) The $P(B)$ curves obtained by following the FCC and FCW protocols coincide for both field orientations and within the full temperature range studied.
To conclude, the above experiment demonstrates that the FLL in CaKFe$_4$As$_4$ sample is well arranged and strongly pinned (at least at $B_{\rm ex}\simeq 11$~mT). The strong pinning in CaKFe$_4$As$_4$ was also reported for $B_{\rm ex}\parallel c$ orientation  in STM experiments,\cite{Fente_PRB_2018} and for both $B_{\rm ex}\parallel ab$ and $B_{\rm ex}\parallel c$ set of data in magnetization studies.\cite{Singh_PRM_2018}

\begin{figure}[htb]
\includegraphics[width=1\linewidth]{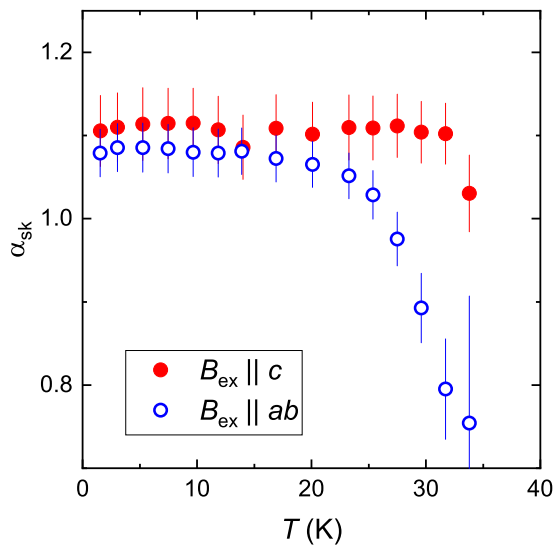}
 \vspace{-1cm}
\caption{The temperature dependence of the skewness parameter $\alpha_{\rm sk}=\langle\Delta B^3 \rangle^{1/3}/\langle\Delta B^2 \rangle^{1/2}_s$ obtained in $B_{\rm ex}\parallel c$ and $B_{\rm ex}\parallel ab$ set of experiments.}
 \label{fig:alpha}
\end{figure}

The TF-$\mu$SR data were analyzed by fitting a three-component expression to the time evolution of the muon-spin polarization (see the Supplemental part, Ref.~\onlinecite{Supplemental_Material}).
The superconducting response of CaKFe$_4$As$_4$ sample was further obtained within the framework of the so-called momentum approach, which includes the calculations of the first moment ($\langle B \rangle$), and the second- ($\langle \Delta B^2 \rangle$) and third-central moments ($\langle \Delta B^3 \rangle$) of the magnetic field distribution function $P(B)$ (see the Supplemental part, Ref.~\onlinecite{Supplemental_Material}, for details).  Following Ref.~\onlinecite{Weber_PRB_1993}, the first moment (the mean field)  scales with the sample magnetization. The second moment (the broadening of the signal), $\langle \Delta B^2 \rangle =\langle \Delta B^2 \rangle_s + \sigma_{nm}^2$, contains contributions from the vortex lattice ($\langle \Delta B^2 \rangle_s$) and the nuclear dipole field [$\sigma_{\rm nm}$, as is obtained from $T>T_{\rm c}$ measurements].  In extreme type-II superconductors ($\lambda\gg\xi$) and for fields $B_{\rm ex}\ll \mu_0 H_{\rm c2}$, $\langle \Delta B^2 \rangle_s \propto \lambda^{-4}$. \cite{Brandt_PRB_1988, Brandt_PRB_2003}
The third moment accounts for the asymmetric shape of $P(B)$, which is described via the skewness parameter $\alpha_{\rm sk}=\langle\Delta B^3 \rangle^{1/3}/\langle\Delta B^2 \rangle^{1/2}_s$.
In the limit of $\lambda\gg\xi$ and for realistic measurement conditions $\alpha_{\rm sk}\simeq 1.2$, for a well arranged triangular vortex lattice.\cite{Aegerter_PRB_1998} It is very sensitive to structural changes of the vortex lattice which may occur as a function of temperature and/or magnetic field.\cite{Lee_PRL_1993, Aegerter_PRB_1998, Blasius_PRL_1999, Menon_PRL_2006, Khasanov_PRL_2008, Heron_PRL_2013}

Figure \ref{fig:alpha} shows temperature dependencies of the skewness parameter $\alpha_{\rm sk}$ measured in $B_{\rm ex}\parallel c$ ($\alpha_{\rm sk}^{\parallel c}$) and $B_{\rm ex}\parallel ab$ ($\alpha_{\rm sk}^{\parallel ab}$) set of experiments. The temperature dependencies of the first moment $\langle B \rangle$ and the second-central moment $\langle \Delta B^2 \rangle$ are presented in the Supplemental part.\cite{Supplemental_Material} Following Fig.~\ref{fig:alpha}, $\alpha_{\rm sk}^{\parallel c}$ stays temperature independent. Only in the close vicinity to $T_{\rm c}$,  $\alpha_{\rm sk}^{\parallel c}$ decreases by approximately 8\% (from $\simeq 1.1$ to $\simeq 1.03$). This, together with the relatively high value of $\alpha_{\rm sk}^{\parallel c}\simeq 1.1$ suggests, that in $B_{\rm ex} \parallel c$ orientation the FLL in CaKFe$_4$As$_4$ is well developed and has little distortions. The presence of well arranged FLL is also confirmed in 'field-shift' experiments [Figs.~\ref{fig:Field-Shift}~(d)--(f)] showing asymmetric $P(B)$'s with all characteristic features attributed to FLL.

The temperature evolution of $\alpha_{\rm sk}^{\parallel ab}$ is, however, different from  $\alpha_{\rm sk}^{\parallel c}(T)$, Fig.~\ref{fig:alpha}.  By increasing temperature, $\alpha_{\rm sk}^{\parallel ab}$ stays almost constant up to $T\simeq 20$~K ($\alpha_{\rm sk}^{\parallel ab}\simeq 1.07$) and it decreases down to $\simeq 0.7$ by approaching $T_{\rm c}$.  In general, the change of $\alpha_{\rm sk}$ as a function of magnetic field and temperature is associated with the vortex lattice melting,\cite{Lee_PRL_1993, Blasius_PRL_1999, Heron_PRL_2013} and/or dimensional crossover from 3D to 2D type of FLL.\cite{Aegerter_PRB_1998, Blasius_PRL_1999} Both processes are thermally activated and caused by increased vortex mobility via loosing the interplanar or intrapalanar FLL correlations.\cite{Blasius_PRL_1999} In both interpretations, however,  the increased vortex mobility leads to decoupling vortices from the pinning centers. This is not the case for CaKFe$_4$As$_4$ studied here. The 'field-shift' experiments, in $B_{\rm ex}\parallel ab$ orientation, suggest of the FLL remaining rigid [at least up to $T\simeq 32$~K, see Fig.~\ref{fig:Field-Shift}~(a)--(c) and the Supplemental part, Ref.~\onlinecite{Supplemental_Material}]. This, therefore, rather indicates that the decrease of  $\alpha_{\rm sk}^{\parallel ab}$ for $T\gtrsim 20$~K is caused by $\gamma_\lambda$ induced rearrangement of the FLL. The change of $\gamma_\lambda$, leading necessarily to the movement of the vortex lines [Fig.~\ref{fig:Gamma}~(a)], enhances the pinning caused FLL distortions and, as a consequence, reduces the value of the skewness parameter $\alpha_{\rm sk}^{\parallel c}$.

\begin{figure}[htb]
\includegraphics[width=1\linewidth]{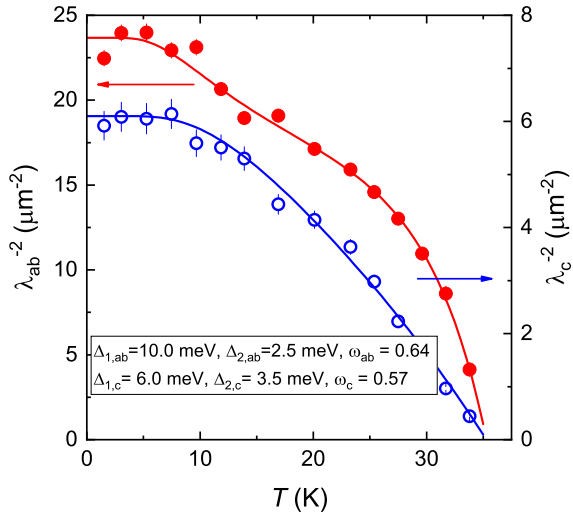}
 \vspace{-1cm}
\caption{Temperature dependencies of the inverse squared in-plane ($\lambda_{ab}^{-2}$) and the out-of-plane ($\lambda_{c}^{-2}$) components ofthe magnetic field penetration depth. Solid lines are fits within the framework of the phenomenological $\alpha-$model.\cite{Khasanov_PRL_2007, Carrington_2003} }
 \label{fig:lambda_ab-lambda_c}
\end{figure}

The $T$ dependence of $\gamma_\lambda$ was further studied via measurements of temperature evolutions of the in-plane ($\lambda_{ab}^{-2}$) and the out-of-plane ($\lambda_{c}^{-2}$) superfluid density components (Fig.~\ref{fig:lambda_ab-lambda_c}). The anisotropy parameter $\gamma_\lambda=\lambda_c/\lambda_{ab}$ is presented in Fig.~\ref{eq:gamma}~(b). Obviously, $\gamma_\lambda$  in CaKFe$_4$As$_4$ is temperature dependent. It is almost constat ($\gamma_\lambda\simeq 1.9$) for $T\lesssim 20$~K and increases up to $\simeq 3.0$ by approaching $T_{\rm c}$. More remarkable is that $\gamma_\lambda(T)$ follows the temperature evolution of the skewness parameter $\alpha_{\rm sk}^{\parallel ab}$. Both, $\gamma_\lambda(T)$ and $\alpha_{\rm sk}^{\parallel ab}$ are {\it constant} for $T\lesssim 20$~K and change almost linearly in $25{\rm ~K}\lesssim T < T_{\rm c}$ region [Figs.~\ref{fig:Gamma}~(b) and \ref{fig:alpha}].  These suggest, that the enhancement of FLL  distortions, seen via decrease of $\alpha_{\rm sk}^{\parallel ab}$ (Fig.~\ref{fig:alpha}) is caused by the vortex lattice rearrangement [Fig.~\ref{fig:Gamma}~(a)], which, in its turn, is initiated by the temperature dependent anisotropy coefficient $\gamma_\lambda$ [Fig.~\ref{fig:Gamma}~(b)].

The comparison of $\gamma_\lambda$ obtained in or study with $\gamma_{H_{\rm c2}}$ reported in Ref.~\onlinecite{Meier_PRB_2016} shows that  these anisotropies coincide with the each other ($\gamma_\lambda=\gamma_{H_{\rm c2}}\simeq 3.0$) only for $T\rightarrow T_{\rm c}$ [Fig.~\ref{fig:Gamma}~(b)], thus resembling the situation in the famous two-gap superconductor MgB$_2$.\cite{Angst_PRL_2002, Angst_Book_2004, Fletcher_PRL_2005} This may suggest that, the range of applicability of AGL to CaKFe$_4$As$_4$ is confined within a narrow range close to $T_{\rm c}$, in full analogy with MgB$_2$, where the AGL description was found to be limited to temperatures less than 2\% away from $T_{\rm c}$.\cite{Golubov_PRB_2003} An alternative approach, based on microscopic Eilenberger calculations, reveals that in two-gap superconductors (as {\it e.g.} MgB$_2$) $\gamma_\lambda$ and $\gamma_{H_{\rm c2}}$ have different temperature dependencies and cross only at $T_{\rm c}$.\cite{Miranovic_JPSJ_2003} Which of these approaches better describe the anisotropic properties of  CaKFe$_4$As$_4$ need further investigation.

The temperature dependencies of  $\lambda_{ab}^{-2}$ and $\lambda_{c}^{-2}$ (Fig.~\ref{fig:lambda_ab-lambda_c}) were further analyzed within the framework of the phenomenological $\alpha-$model by decomposing $\lambda^{-2}(T)$ into two components:\cite{Khasanov_PRL_2007,Padamsee_JLTP_1973, Carrington_2003}
\begin{equation}
\frac{\lambda^{-2}(T)}{\lambda^{-2}(0)}=
\omega\; \frac{\lambda^{-2}(T,
\Delta_{1})}{\lambda^{-2}(0,\Delta_{1})}+(1-\omega)\;
\frac{\lambda^{-2}(T, \Delta_{2})}{\lambda^{-2}(0,\Delta_{2})}.
 \label{eq:two-gap}
\end{equation}
Here $\lambda(0)$ is the value of the penetration depth at $T=0$, $\Delta_i$ is the zero-temperature value of $i-$th gap ($i=1$ or 2) and $\omega$ ($0\leq\omega\leq1$) is the weight of  the larger gap to  $\lambda^{-2}$. Each component in Eq.~\ref{eq:two-gap} was further evaluated by $\lambda^{-2}(T,\Delta)/\lambda^{-2}(0,\Delta)= 1 +2\int_{\Delta(T)}^{\infty}\left(\partial f / \partial E \right)E /\sqrt{E^2-\Delta(T)^2} \;
dE$.\cite{Tinkham_75}
Here $f=[1+\exp(E/k_{\rm B}T)]^{-1}$ is  the Fermi function and $\Delta(T)=\tanh \{1.82[1.018(1/t-1)]^{0.51} \}$.\cite{Carrington_2003} Note that this equation is valid within the clean-limit, which is indeed the case for CaKFe$_4$As$_4$.\cite{Meier_PRB_2016}
Solid lines in Fig.~\ref{fig:lambda_ab-lambda_c} are fits of Eq.~\ref{eq:two-gap} to $\lambda_{ab}^{-2}(T)$ and $\lambda_{c}^{-2}(T)$ data. The fit parameters are: $\lambda_{ab}^{-2}(0)=23.4$~($\mu$m$^{-2}$), $\Delta_{1,ab}=10.0$~meV, $\Delta_{2,ab}=2.5$~mev $\omega_{ab}=0.64$, and $\lambda_{c}^{-2}(0)=6.1$~($\mu$m$^{-2}$), $\Delta_{1,c}=6.0$~meV, $\Delta_{2,c}=3.5$~meV, $\omega_c=0.57$ for $\lambda_{ab}^{-2}(T)$ and $\lambda_{c}^{-2}(T)$ data sets, respectively.

From the results of two-gap model fit to $\lambda_{ab}^{-2}(T)$ and $\lambda_{c}^{-2}(T)$, three important points emerge:
(i) The 'theoretical' two-gap $\gamma_\lambda(T)$ curve stays in a good agreement with the experimental data [see Fig.~\ref{fig:Gamma}~(b)].
(ii) The density functional theory calculations suggest that in CaKFe$_4$As$_4$ ten bands (6 hole- and 4 electron-like bands) having different zero-temperature gap values cross the Fermi level.\cite{Mou_PRL_2016, Lochner_PRB_2017} Recent angle resolved photoemission (ARPES),\cite{Mou_PRL_2016} as well as the combination of ARPES and $\mu$SR experiments,\cite{Khasanov_PRB_2018} confirm this statement. This suggests, on the one hand, that the two-gap approach described by Eq.~\ref{eq:two-gap} is oversimplified and contributions of all ten bands need to be included. Such analysis requires, however, an exact knowledge of the electronic band structure.\cite{Khasanov_PRB_2018, Evtushinsky_NJP_2009} On the other hand, similar two-gap approaches were previously used in Refs.~\onlinecite{Meier_PRB_2016, Cho_PRB_2017, Biswas_PRB_2017} and allowed to describe satisfactorily the in-plane superfluid density data of  CaKFe$_4$As$_4$.
(iii) The difference in parameters obtained from the two-gap fit to $\lambda_{ab}^{-2}(T)$ and $\lambda_{c}^{-2}(T)$ are most probably caused by {\it anisotropic} contributions of various bands to the superfluid density. In such case $\Delta_1$ and $\Delta_2$ refer to averaged gaps for series of bands with 'big' and 'small' gap values, respectively (see {\it e.g.} Ref.~\onlinecite{Lochner_PRB_2017}), while $\omega$ correspond to the averaged weight of 'big' gaps to $\lambda^{-2}(0)$.

To conclude, the magnetic penetration depth anisotropy $\gamma_\lambda=\lambda_{c}/\lambda_{ab}$ in a CaKFe$_4$As$_4$ single crystal sample was studied by means of the muon-spin rotation. $\gamma_\lambda$ is temperature independent for $T\lesssim 20$~K ($\gamma_\lambda\simeq 1.9$) and it increases up to $\gamma_\lambda\simeq 3.0$ for  $T \rightarrow T_{\rm c}$. The change of $\gamma_\lambda$ induces the flux line lattice rearrangement, which is clearly detected via the enhanced distortions of FLL $\mu$SR response. The anisotropic properties of $\lambda$ are explained by the multi-gap nature of superconductivity in CKaFe$_4$As$_4$ and  caused by anisotropic contributions of various band to the in-plane and the out-of-plane components of the superfluid density.

The work was performed at the Swiss Muon Source (S$\mu$S), Paul Scherrer Institute (PSI, Switzerland). Authors acknowledge helpful discussions with V.G. Kogan.  WRM was funded by the Gordon and Betty Moore Foundation EPiQS Initiative through Grant GBMF4411. Work at Ames Laboratory was supported by the U.S. Department of Energy, Office of Science, Basic Energy Sciences, Materials Science and Engineering Division. Ames Laboratory is operated for the U.S. DOE by Iowa State University under Contract No. DE-AC02-07CH11358.

\end{document}